\documentclass[apj]{emulateapj}
\usepackage{graphicx}
\usepackage{hyperref}
\usepackage{amsmath}
\usepackage{epstopdf}
\newcommand{\kms}[0]{km\,s$^{-1}$}
\newcommand{\glimpse}[0]{{\sc glimpse} }
\newcommand{\vla}[0]{{\sc vla} }
\newcommand{\magpis}[0]{{\sc magpis} }
\newcommand{\citepy}[1]{\citeauthor{#1} (\citeyear{#1})}
\newcommand{\hii}[0]{H{\sc ii} }
\newcommand{\jcmt}[0]{{\sc jcmt} }
\graphicspath{{figs/}}
\begin{document}
\slugcomment{accepted for publication into ApJ}
\shorttitle{Molecular Rings and the Thickness of Star-Forming Clouds}
\title{Molecular Rings around Interstellar Bubbles and the \\ Thickness of Star-Forming Clouds}
\shortauthors{Beaumont \& Williams}
\author{Christopher N. Beaumont \& Jonathan P. Williams}
\affil{Institute for Astronomy, University of Hawaii \\
2680 Woodlawn Drive, Honolulu, HI 96822}
\email{beaumont,jpw@ifa.hawaii.edu}

\begin{abstract}

The winds and radiation from massive stars clear out large cavities in
the interstellar medium. These bubbles, as they have been called,
impact their surrounding molecular clouds and may influence the
formation of stars therein. Here we present JCMT observations of the
$J=3-2$ line of CO in 43 bubbles identified with Spitzer Space
Telescope observations. These spectroscopic data reveal the
three-dimensional structure of the bubbles.
In particular, we show that the cold gas lies in a
ring, not a sphere, around the
bubbles indicating that the parent molecular clouds
are flattened with a typical thickness of a few parsecs.
We also mapped 7 bubbles in the $J=4-3$ line of HCO$^+$
and find that the column densities inferred from the
CO and HCO$^+$ line intensities are below that necessary 
for ``collect and collapse'' models of induced star formation. We hypothesize that
the flattened molecular clouds are not greatly compressed by expanding
shock fronts, which may hinder the formation of new stars.

\end{abstract}
\keywords{ISM: kinematics and dynamics --- stars: formation --- (ISM:) \hii regions }  
\maketitle

\section{Introduction}
\label{sec:intro}

Massive stars exhibit strong winds and radiation fields which clear
out parsec-scale cavities in the interstellar medium (ISM) \citep{Castor75, Weaver77}.
At the boundaries of these cavities (or bubbles) lie the
material displaced by the propagating wind or ionization front. Many
have argued that such a dynamic process, which alters the physical
environment of molecular clouds, might trigger the formation of a
new generation of stars. Various mechanisms are reviewed by \citepy{Elmegreen98}.

Bubble surfaces are illuminated by the central stars and shine brightly
in the infrared. A by-product of the Spitzer Galactic Legacy Infrared
Mid-Plane Survey Extraordinaire ({\sc glimpse}; \citealt{Benjamin03})
was the identification of nearly 600 bubbles \citep{Churchwell06, Churchwell07}.
Because of Spitzer's tremendous increase in
sensitivity and resolution over previous infrared observatories, the
\glimpse bubble catalogs represent a dataset of interstellar
bubbles unprecedented in size and detail.  Simultaneously,
sub-millimeter astronomy has enjoyed rapid technological
improvements. The newly-commissioned HARP heterodyne receiver
array on the James Clerk Maxwell Telescope (JCMT) provides the
ability to efficiently map large portions of the submillimeter sky at high
spatial and velocity resolution \citep{Smith08}. This
permits a followup study of the molecular gas in a sizable
subset of the original bubble catalogs.

In this paper we present HARP observations of the $J=3-2$ CO line
toward 43 bubbles in the \citepy{Churchwell06} catalog.
The data reveal the three-dimensional structure of the bubbles
and their interaction with the ambient molecular cloud.
We also observed the $J=4-3$ HCO$^+$ line toward 7 bubbles
to determine the amount of dense gas in their surroundings.
These data are analyzed together with the \glimpse catalogs,
and with 20\,cm radio continuum images of the ionized gas,
to quantify the degree of star formation in and around the bubbles.

We find that the molecular gas tends to lie in rings and conclude
that the bubbles quickly break out of their molecular surroundings.
This, in turn, suggests that molecular clouds are oblate with
one dimension significantly thinner than the other two.
We investigate the extent of cloud compression and star formation in these regions,
and posit that triggered star formation may be hindered due
to the central stars' rapid breakout from their molecular confines.

In \S\ref{sec:observations} we describe our observations, and introduce the supplementary infrared and radio datasets
we have utilized. We present and discuss our analysis of these data in \S\ref{sec:results}. In \S\ref{sec:discussion}, 
we interpret our findings in the context of current theories about molecular cloud structure and star formation. We conclude in
\S\ref{sec:conclusion}.

\section{Observations}
\label{sec:observations}

HARP is a $4\times 4$ heterodyne array on the JCMT
designed for large scale mapping \citep{Smith08}. Under typical
atmospheric conditions on Mauna Kea, the HARP array can map 345 GHz line emission 
to a noise level of T$_{\rm rms} = 0.3$\, K at a rate of 
100 arcmin$^{2}$ hour$^{-1}$.

We selected targets at Galactic longitude $l>10^\circ$,
and with angular radii $2' < \theta < 10'$ from the Churchwell
catalogs of 592 bubbles. These constraints are based on the
practical considerations of being observable from Hawaii in the
summer and being well resolved but not so large as to require a
prohibitively long time to map. 
As the bubbles span a wide range in distance and size,
there is little correlation between the angular and intrinsic size.
Hence, our target list is a fairly uniform sampling of the original catalog.

Our observations were carried out over ten nights in 2008 July-August.
Data were acquired in raster observing mode, using a nearby
reference position that was first verified to be free of emission.
We fit and removed a first order baseline to the spectral data away from the line, and then 
gridded and coadded the individual raster pointings. Finally, we binned the coadded spectral
cubes to 0.2 km\, s$^{-1}$, and smoothed the data with a $6''$ Gaussian kernel. The final spatial resolution
of each cube is $16 ''$. All data reduction was performed using the STARLINK \footnote{http://starlink.jach.hawaii.edu/starlink}
package.
To convert the intensities to radiation temperatures, we adopted 
a main beam efficiency of 70\%.
In total, we observed 45 rings in the CO 3--2 transition,
and 7 in HCO$^+$ 4--3. The reduced data have typical rms values of $T_{\rm rms} = 0.3 $K.

To supplement our sub-mm data, we retrieved image and catalog data of
these regions from the Spitzer \glimpse and \vla {\sc
magpis} surveys \citep{Helfand06}. The infrared data provide high resolution images of the bubbles, 
and help identify bubble features in the sub-mm data. Furthermore, we use the \glimpse data 
to identify and study young stellar objects (YSOs) in these regions. 

The \vla {\sc magpis} survey is a multi-configuration \vla
survey of cm-emission in the galactic plane \citep{Helfand06}. We use these data to 
identify \hii regions associated with our sample. The intensity of \hii regions provides
an estimate of the ionizing photon flux in the region, which in turn constrains the mass of the central star(s)
driving bubble expansion.

\section {Results}
\label{sec:results}
\subsection {Molecular rings}

CO is the most abundant molecule with a low-energy transition
in the ISM. The $J=3-2$ transition is an excellent tracer of the
cool, 20-50K, moderately dense, $n_{\rm H_2}\sim 10^{3-4}\,{\rm cm}^{-3}$,
gas around the bubbles.  The emission from this line traces
the ambient molecular cloud around the bubble shells, and is particularly 
strong at their interface.
HCO$^+$ is moderately abundant but has strong emission in
gas with densities $n_{\rm H_2}>10^4\,{\rm cm}^{-3}$.
It is a useful signpost of star-forming and potentially pre-stellar
gas.

\begin{figure}
\includegraphics[height=3.5in, trim= 30mm 0mm 10mm 0mm, angle=90]{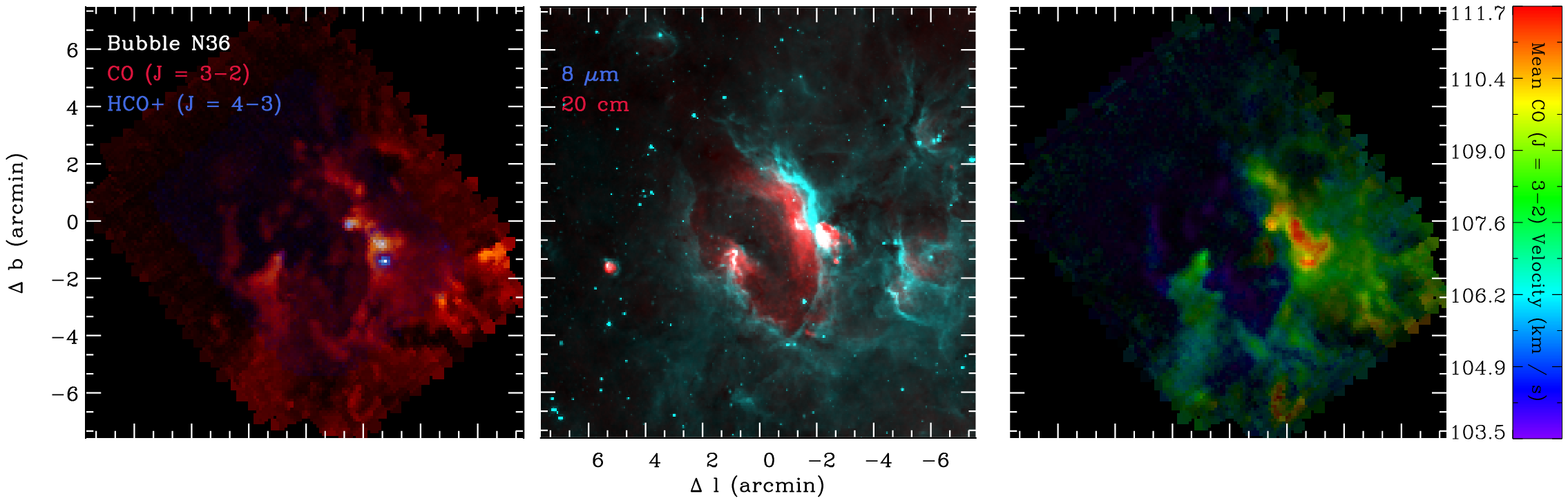}
\includegraphics[height=3.5in, trim= 10mm 0mm 20mm 0mm, angle=90]{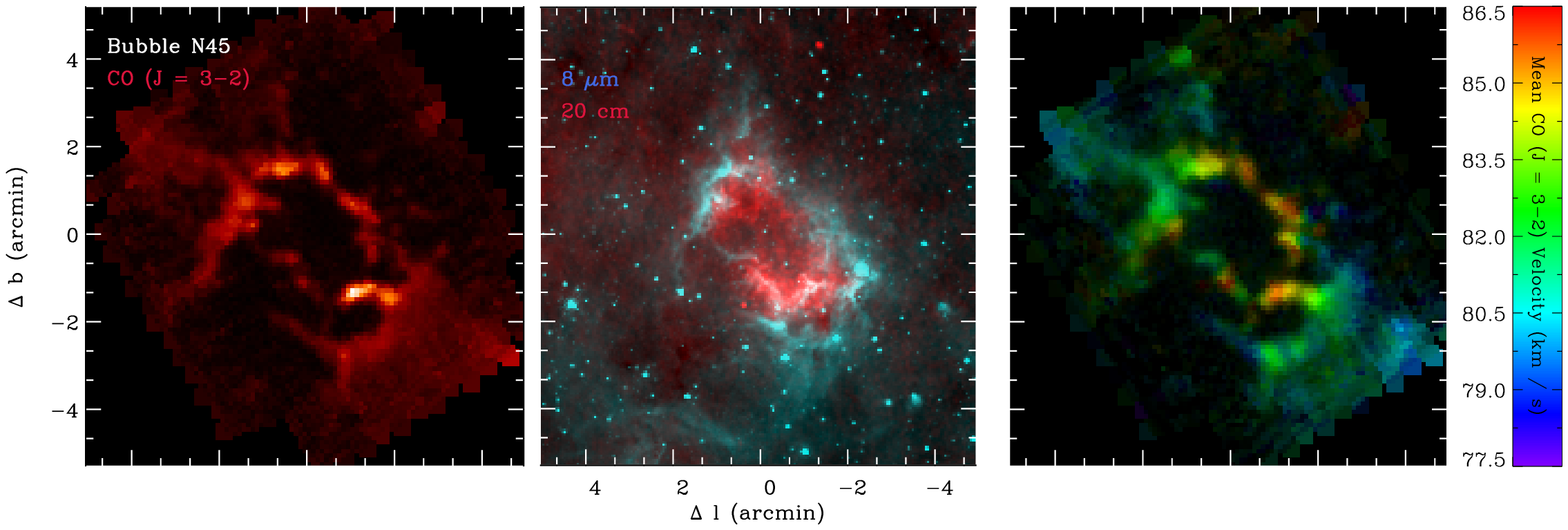}
\caption{Examples from our dataset. The left figures show the peak intensity of the CO and, when observed, HCO$^+$ emission. 
The center figures show Spitzer $8 \mu$m emission in blue, and \vla 20 cm emission in red. The rightmost figures show the 
first moment of the CO emission. The rings shown
are N36 (top) and N45 (bottom). Equivalent figures for our
entire sample can be found in the electronic version of this article.}
\label{fig:bubbleExample}
\end{figure}

We detected CO emission toward 43 of the 45 bubbles that we observed
and HCO$^+$ in 6 of the 7 followup targets.
Images of the molecular emission in relation to the Spitzer $8\,\mu$m
and, where available, 20\,cm data are shown in Figure~\ref{fig:bubbleExample}.
Figures for all 43 bubbles in our sample can be found in the electronic version 
of this article. 

There is generally a very good correspondence between the
CO and Spitzer maps; because the interiors of bubbles are well-evacuated, 
they are clearly defined in the sub-mm data cubes. There are occasional 
departures between the infrared and sub-mm morphologies. For example,
filamentary structures can be seen in many of the sub-mm data, but do not
emit prominently in the infrared. These structures extend both out of 
(e.g., bubbles N37, N45, N90) and into
(e.g., N29, N27, N44, N49) the bubbles themselves. 

The CO 3-2 line is readily excited by collisions in molecular clouds,
and very easily becomes optically thick. Because of the CO's opacity,
the \jcmt data cubes show the surfaces (in position-velocity space) of
molecular clouds. The filamentary structures in these data illustrate the inhomogeneity
of molecular clouds on parsec-scales. These inhomogeneities are not as well probed 
at mid-infrared wavelengths. The emission at $8\, \mu$m is dominated by
UV-excited PAH emission, and is expected to be optically thin 
\citep{Churchwell06, Watson08}. The more uniform nature of bubble emission at
these wavelengths most likely reflects the (approximately) spherically symmetric
UV radiation field produced by the bubble-blowing stars. 

The most striking feature of the CO data is the rarity of emission towards 
the center of the bubbles. If bubbles are 2-dimensional projections
of spherical shells, we would expect to see the front and back faces of these
shells at blueshifted and redshifted velocities. Of the 43 bubbles we have studied,
we find no convincing examples of such a structure. 

Three different lines of evidence suggest that the molecular data are inconsistent
with a spherical shell morphology. First is the contrast ratio between the bubble shells
and bubble interiors. In the simplest case of optically thin emission, the integrated intensity
scales linearly with the line-of-sight path length
through the shell, $l$, as a function of impact parameter $b$,
\begin{equation}
\label{eq:othin}
l = \left\{
\begin{array}{l l}
   2R[(1 + \Delta R/R)^{2} - (b/R)^{2}]^{\frac{1}{2}}&  - 2R[1 - (b/R)^{2}]^{\frac{1}{2}} \\
    & \quad \mbox{$b < R $}  \\
	\\
   2R[(1 + \Delta R/R)^{2} - (b/R)^{2}]^{\frac{1}{2}} & \quad \mbox{$R \leq b < R + \Delta R$}\\
   \\
   0 &\quad \mbox{$b \geq R + \Delta R$}
\end{array} \right.
\end{equation}
where the bubble has an inner radius $R$ and thickness $\Delta R$. \citepy{Watson08} have
used this equation to describe the $8 \mu$m radial intensity profile for bubble N49. 

Compared to $8 \mu$m emission, CO provides a stronger test of this equation for two reasons. First, the 
added velocity information provides some ability to separate bubble emission from unassociated
foreground and background material. Second, the expected opacity of CO serves to decrease
the edge-to-center contrast ratio, making the bubble fronts and backs easier to detect if present 
(the extreme profile is a completely opaque disk with no radial intensity variation). 

\begin{figure}
\includegraphics[height=3.5in, angle=90]{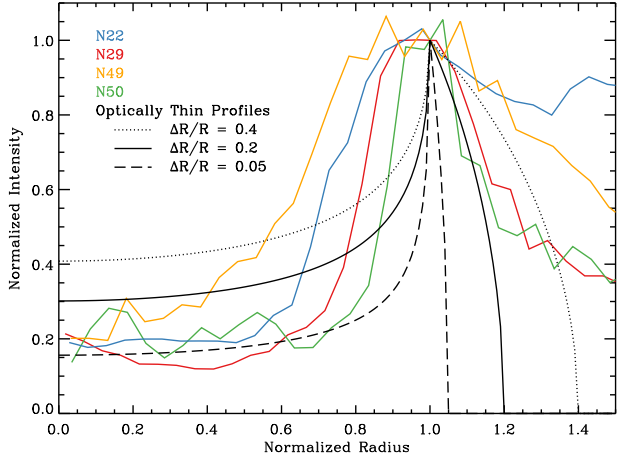}
\caption{Plot of the azimuthally averaged radial profile of four rings, compared to
profiles for optically thin spherical shells. The data are inconsistent with these profiles, which have overly thin outer
edges or low edge-to-center contrast ratios. This strongly disfavors the spherical shell model for bubble morphologies, and suggests instead that
these bubbles lack fronts and backs.}
\label{fig:contrast}
\end{figure}

Figure \ref{fig:contrast} plots Equation \ref{eq:othin} for three different
values of $\Delta R / R$. The azimuthally averaged radial
profiles of four well defined bubbles are also plotted on this
Figure. The Figure illustrates that the contrast ratio between
the CO shells and interiors is problematically high. To reproduce such high contrasts, very thin shells must be invoked.
However, the shell thicknesses are resolved in several of the larger objects in our sample,
and rule out the very-thin shell hypothesis. Furthermore, as mentioned above, treating the CO emission
as optically thin is an unrealistic assumption, and its opacity makes the dearth of emission interior to
bubble shells even harder to reconcile to a spherical shell morphology.

The intensity of emission surrounding bubbles provides the second line of evidence against the spherical shell hypothesis. 
Most of the objects in our sample are found to lie interior to, or on the edge of, regions of extended molecular emission. 
If bubbles are embedded in molecular material, then these clouds should contribute foreground and background
emission inside bubble shells.
However, the CO interiors of many bubbles are actually \textit{darker} than their immediate exteriors (see, 
e.g., N21, N22, N29, N49, N74, N80). We note that even the $8\, \mu $m emission occasionally displays this signature
(bubbles from this study include N90, N94 and, to a lesser extent, N36 and N37).

\begin{figure}
\includegraphics[width=3.5in, angle=0]{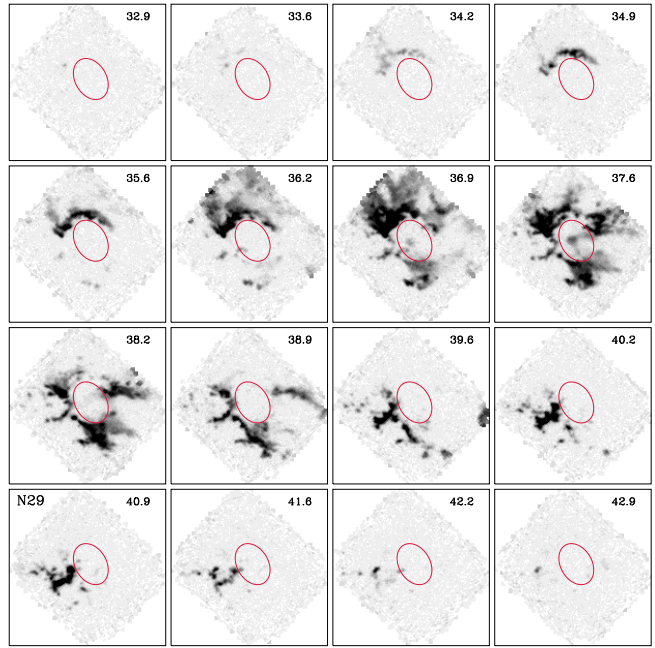}
\caption{Example channel map for bubble N29. The radial velocity at each slice is listed in km\, s$^{-1}$ in the upper right corner. 
An ellipse is drawn in for reference, to mark the approximate boundary of the interior. 
Note that the bubble interior is largely devoid of emission. The few structures that do appear inside the shell (for example, 
at a radial velocity of 37.6~km~s$^{-1}$) are not red- or blueshifted, as would be expected if N29 was an expanding sphere.
Note also that the extended emission to the north of the bubble does not extend into the interior of N29. This points to 
an absence of ambient molecular material in front of or behind the bubble.}
\label{fig:channelmap}
\end{figure}

Finally, the expanding spherical shell hypothesis makes predictions about the velocity structure of bubbles which
are not borne out by the CO data. An expanding spherical shell would manifest itself as a
connected, coherent structure in a data cube. Specifically, as one
steps through velocity space, a sphere would appear as a blue-shifted point (the
bubble front) expanding into a ring (the midsection), and then
contracting back to a red-shifted point (the back). A channel map for bubble N29 is displayed in Figure \ref{fig:channelmap}. 
The occasional filamentary structures seen interior to bubbles do not display the systematic velocity shifts as expected, but are instead moving
at roughly the same radial velocity as the edges. These filaments are either unaffiliated foreground or
background structures (consistent with the observation that they are absent from the Spitzer data), or they
imply that the expansion of these regions has stalled. 

\begin{figure}
\includegraphics[width=3.5in]{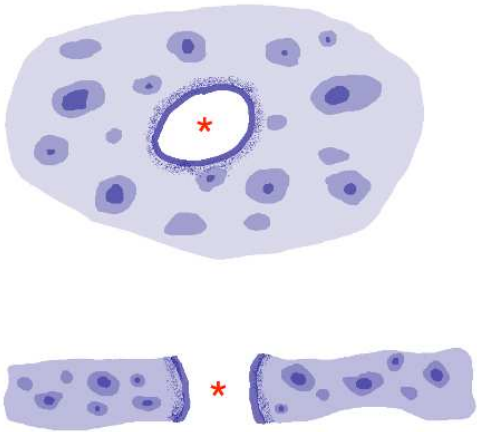}
\caption{A schematic of an interstellar ring, where a central group of O and B stars clear out a cavity in a flattened molecular cloud.  
The top panel shows a face on view of an approximately-circular cavity of diameter a few parsecs. None, or very little, CO emission
is seen towards the cavity, whereas the molecular cloud extends for tens of parsecs in the plane of the sky. We interpret our finding 
as indicating that clouds have line-of-sight depths of a few parsecs, and that bubbles rapidly break out of their host clouds 
(illustrated in the lower panel). 
In this scenario, the interaction region between the star and cloud is smaller than it would be than for a more homogeneous cloud geometry. 
}
\label{fig:schematic}
\end{figure}

In light of this evidence, the most natural interpretation to draw from the observations
is that the fronts and backs of bubbles are missing,
and we are instead observing only the bubble midsection.
A corollary is that the molecular clouds in which bubbles are
embedded are oblate with thicknesses
of a few parsecs, comparable to typical ring diameters (Figure \ref{fig:schematic}).
Expanding bubbles break out of the cloud along this axis
and we see a ring of CO emission, approximately circular if
the flattened axis is along our line-of-sight, otherwise
elliptical. If viewed edge-on, the bubbles would not be identifiable
as such either in our images or those from Spitzer, but might
perhaps be classified as filamentary structures or Infrared Dark Clouds \citep{Simon06, Jackson08}.

The relationship between the ring diameter and its parent cloud's thickness depends on
the mechanism driving the ring's expansion. If the stars powering the bubble produce
stellar winds of sufficient strength, then the bubble expansion is driven by direct momentum
transfer from the wind into the ambient material \citep{Castor75}. In this scenario, the ring's expansion would continue after 
it breaches the flattened cloud, and the observed ring diameter places an upper limit
on the thickness of the surrounding material. Alternatively, if the stars power an \hii region
but do not have powerful winds, then a shock from the pressure-driven expansion of the \hii region surrounds the 
stellar wind shock front \citep{Freyer03}.  Rings in this case trace the boundary of the \hii shock and, once the bubble
breaches the cloud top and bottom, the pressure would drop and expansion would stall. In this
case, the ring diameter is roughly equal to the cloud thickness. As we discuss in Section \ref{sec:central sources}, 
both classes of bubbles are present in this sample.

The ring-like morphology of cold molecular gas encircling spherically
expanding structures appears to be commonplace: it is present not only
in the data presented here, but also in many other well studied
examples of large \hii regions \citep{Williams95, Koenig08, Zavagno06, Deharveng09}. 
However, the implications of this
observational signature have not generally been commented
upon. Furthermore, these bubbles are smaller than typical optically visible \hii regions,
and thus place stronger constraints on their parent clouds' thicknesses.

\subsection {Ring properties}
\label{sec:RingProp}

While the molecular rings can often be easily identified by eye in a
CO data cube, it is harder to rigorously define where the bubble interaction zone
ends and the ambient molecular cloud begins. Here we describe our
process of isolating bubble features in the data.

\begin{figure}
\includegraphics[width=3.5in, angle=0]{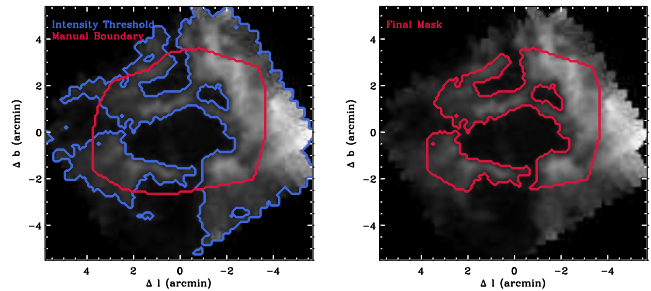}
\caption{An illustration for how we defined each ring's boundary in the CO data cubes. 
The red and blue lines on the left trace, respectively, the intensity-thresholded and manually-defined boundaries for bubble N22; 
both boundaries are needed to effectively define the inner and outer ring boundaries. 
The figure on the right is the intersection of these two regions, and defines which pixels we consider to be part of this bubble.}
\label{fig:thresh}
\end{figure}

We first cropped the data in velocity to include only those channels
with matching features in the Spitzer images. In most cases, the inner
ring edge stands out in the integrated maps. Thus, the inner boundary
is defined by an integrated intensity threshold, chosen manually for
each ring. While some outer ring boundaries stand out against dark
backgrounds in the integrated maps, rings on the edges of extended
cloud structures do not. As a compromise, we define each ring's outer boundary as the intersection of 
the intensity threshold and a manual boundary. An example of this process is illustrated in
Figure~\ref{fig:thresh}; the red line on the left traces the boundary
defined by hand, while the blue line traces the intensity threshold;
the figure on the right shows the intersection of these two regions,
and serves to define which pixels are associated with this ring.

The ring properties for our 43 CO detections are listed in
Table~\ref{tab:main}. Near kinematic distances are calculated using the galactic rotation
model of \citepy{Brand93}, for $R_{\sun} = 8.8$ kpc and
$V_{\sun} = 275$ km/s \citep{Reid09}. Kinematic distance
uncertainties are dominated by peculiar motions. Based on the data
from \citepy{Reid09}, we estimate this uncertainty to be about 30\%, and use this value for the 
distance errors reported in Table~\ref{tab:main}. 

\citepy{Churchwell06} make the argument that bubbles are more
likely located at their near kinematic distances, since interstellar
extinction and confusion with other diffuse emission structures
would tend to obscure objects on the far side of the
Galactic disk. Following this assumption, we convert the Spitzer
angular shell radii and thicknesses to linear measurements, and
include them in Table \ref{tab:main}.

Seventeen of the bubbles in our CO survey are associated with the \hii regions in \citepy{Paladini03}. 
Our independent velocity measurements agree with the recombination line velocities in that data set, 
strengthening the argument -- made previously on statistical
grounds -- that features seen in the infrared and submillimeter are
physically associated with these ionized regions. Furthermore, we
provide 26 new kinematic distance measurements.  We note that submillimeter data
have the advantage of providing distance estimates to rings without
detected \hii regions. 

\begin{center}
\begin{deluxetable*}{lrrrrrrrr}[!hb]
\tablecolumns{9}
\tablewidth{0in}
\tabletypesize{\scriptsize}
\tablecaption{Bubble properties \label{tab:main}}
\tablehead{
\colhead{Bubble\tablenotemark{a}} &  \colhead{$l$}  &  \colhead{$b$}  & \colhead{$v_{\rm CO}$}  & \colhead{$\sigma_v$} & \colhead{$d_{\rm near}$} & \colhead{$\sigma_d$} & \colhead{$R$\tablenotemark{b}} & \colhead{$\Delta R$\tablenotemark{b}}\\
\colhead{} & \colhead{($^\circ$)} & \colhead{($^\circ$)} & \colhead{(km\,s$^{-1}$)} & \colhead{(km\,s$^{-1}$)} & \colhead{(kpc)} & \colhead{(kpc)} & \colhead{(pc)} & \colhead{(pc)}} 
\startdata

  N5  &  12.46 &  -1.12 & 39.5 & 2.4 & 3.7 & 1.1 & 3.49 & 0.71\\
 N10  &  13.20 &   0.06 & 50.2 & 4.1 & 4.1 & 1.2 & 1.57 & 0.42\\
 N14  &  13.99 &  -0.13 & 40.3 & 2.8 & 3.5 & 1.1 & 1.24 & 0.39\\
 N15  &  15.01 &  -0.61 & 19.7 & 5.2 & 1.9 & 0.6 & 0.83 & 0.25\\
 N16  &  14.98 &   0.06 & 25.2 & 3.0 & 2.4 & 0.7 & 1.73 & 0.44\\
 N20  &  17.92 &  -0.67 & 42.1 & 3.3 & 3.2 & 0.9 & 1.04 & 0.25\\
 N21  &  18.19 &  -0.39 & 48.1 & 3.1 & 3.4 & 1.0 & 2.14 & 0.49\\
 N22  &  18.26 &  -0.30 & 51.3 & 2.1 & 3.6 & 1.1 & 1.77 & 0.45\\
 N27  &  19.81 &   0.02 & 57.6 & 2.6 & 3.8 & 1.1 & 1.19 & 0.24\\
 N29  &  23.06 &   0.56 & 37.9 & 2.1 & 2.5 & 0.8 & 1.98 & 0.59\\
 N30  &  23.11 &   0.58 & 38.7 & 2.6 & 2.6 & 0.8 & 0.81 & 0.26\\
 N34  &  24.30 &  -0.17 & 101.0 & 1.8 & 5.2 & 1.5 & 1.65 & 0.41\\
 N35  &  24.50 &   0.24 & 115.0 & 6.1 & 5.7 & 1.7 & 5.41 & 1.54\\
 N36  &  24.84 &   0.11 & 109.0 & 5.2 & 5.5 & 1.6 & 4.51 & 1.39\\
 N37  &  25.30 &   0.30 & 40.8 & 1.6 & 2.6 & 0.8 & 1.34 & 0.37\\
 N39  &  25.36 &  -0.15 & 63.8 & 3.8 & 3.7 & 1.1 & 2.14 & 0.70\\
 N40  &  25.37 &  -0.36 & 54.8 & 3.1 & 3.3 & 1.0 & 1.18 & 0.36\\
 N44  &  26.83 &   0.38 & 81.1 & 1.8 & 4.3 & 1.3 & 1.41 & 0.30\\
 N45  &  26.99 &  -0.05 & 83.2 & 3.4 & 4.4 & 1.3 & 1.88 & 0.68\\
 N46  &  27.31 &  -0.12 & 92.6 & 2.8 & 4.8 & 1.4 & 1.93 & 0.67\\
 N47  &  28.03 &  -0.03 & 99.2 & 4.4 & 5.1 & 1.5 & 3.32 & 0.76\\
 N49  &  28.83 &  -0.23 & 87.5 & 3.1 & 4.6 & 1.4 & 1.77 & 0.43\\
 N50  &  28.99 &   0.10 & 70.2 & 2.4 & 3.8 & 1.1 & 1.88 & 0.45\\
 N51  &  29.15 &  -0.26 & 94.7 & 3.0 & 4.9 & 1.5 & 2.79 & 0.60\\
 N52  &  30.74 &  -0.02 & 94.1 & 9.6 & 4.8 & 1.5 & 2.85 & 0.87\\
 N53  &  31.16 &  -0.14 & 41.6 & 2.6 & 2.4 & 0.7 & 0.59 & 0.14\\
 N54  &  31.16 &   0.30 & 102.0 & 4.8 & 5.2 & 1.6 & 2.69 & 0.57\\
 N56  &  32.58 &   0.00 & 100.0 & 2.8 & 5.2 & 1.6 & 1.69 & 0.42\\
 N61  &  34.15 &   0.14 & 57.3 & 3.5 & 3.1 & 0.9 & 3.15 & 0.51\\
 N62  &  34.33 &   0.21 & 56.2 & 4.2 & 3.1 & 0.9 & 1.37 & 0.30\\
 N65  &  35.02 &   0.33 & 52.5 & 2.6 & 2.9 & 0.9 & 1.81 & 0.46\\
 N74  &  38.93 &  -0.40 & 40.4 & 2.6 & 2.3 & 0.7 & 0.92 & 0.31\\
 N77  &  40.42 &  -0.05 & 68.8 & 2.0 & 3.9 & 1.2 & 1.46 & 0.36\\
 N79  &  41.52 &   0.03 & 14.1 & 3.1 & 0.7 & 0.2 & 0.26 & 0.06\\
 N80  &  41.93 &   0.03 & 17.7 & 2.8 & 0.9 & 0.3 & 0.48 & 0.11\\
 N82  &  42.11 &  -0.62 & 66.5 & 1.8 & 3.8 & 1.1 & 1.86 & 0.49\\
 N84  &  42.83 &  -0.16 & 61.2 & 1.2 & 3.5 & 1.1 & 1.18 & 0.31\\
 N90  &  43.76 &   0.09 & 69.4 & 2.3 & 4.1 & 1.2 & 2.08 & 0.50\\
 N92  &  44.33 &  -0.83 & 62.0 & 2.6 & 3.7 & 1.1 & 2.09 & 0.57\\
 N94  &  44.81 &  -0.48 & 47.0 & 2.1 & 2.7 & 0.8 & 2.95 & 0.91\\
N120  &  55.27 &   0.25 & 29.3 & 2.6 & 1.9 & 0.6 & 0.76 & 0.17\\
N130  &  62.37 &  -0.54 & 0.4 & 0.3 & 8.4 & 2.5 & 2.57 & 0.64\\
N133  &  63.15 &   0.44 & 21.1 & 3.1 & 1.6 & 0.5 & 0.76 & 0.18\\

%\nodata 

\enddata
\tablenotetext{1}{from Churchwell et al. (2006)}
\tablenotetext{2}{assuming near kinematic distance}
\end{deluxetable*}
\end{center}

\subsection {Constraints on the central sources}
\label{sec:central sources}

The stars that create bubbles also ionize the material in their immediate vicinity, which
glows via recombination and free-free processes. The \vla \magpis survey is sensitive to this emission,
and can constrain the types of stars responsible for the ionization. The \magpis survey overlaps 40 bubbles in our sample.

While free-free emission is clearly evident in the majority of our sample 
(see Figure \ref{fig:bubbleExample}), there are several bubbles for which radio emission is faint or hidden
in the background. For both bright and faint \hii regions, we derive a flux estimate in the following way. We use aperture photometry to compute
the integrated signal within each bubble, corrected for any nonzero background level. We refer to this signal measurement as $D$, because it is derived
from the data. Because we find that the background noise in the
\magpis maps is well described by a Gaussian distribution, the error on a spatially integrated flux measurement is also distributed as a Gaussian,
with standard deviation $\sigma = \sigma_0 \sqrt{\Omega / \Omega_B}$. Here, $\sigma_0$ is the standard deviation of the background noise, $\Omega$ is the solid angle
subtended by the object of interest, and $\Omega_B$ is the solid angle subtended by the antenna beam. Using Bayes' Theorem, the posterior distribution for 
a source's true integrated flux F is then given by
\begin{equation}
\label{eq:bayes}
P(F | D) = \frac{P(D | F)P(F)}{\int_{-\infty}^\infty\,P(D | F)P(F)\,dF}
\end{equation}

\noindent Where $P(D|F)$ is the likelihood measuring D when the true flux is $F$; it is simply a Gaussian centered on $F$, with a width of $\sigma$, evaluated at $D$. 
The function $P(F)$ is the prior probability distribution for F before any data were taken; we define this function to be zero
for $F<0$ (since no real source can produce a negative flux), and constant otherwise. The actual value of the constant is irrelevant, since it cancels with
the denominator.

The benefit of deriving flux estimates from Equation \ref{eq:bayes} is twofold. First, the prior $P(F)$ naturally incorporates the the physical 
restriction that the true flux be non-negative. Second, it makes no distinction between detections and non-detections, and is equally valid
in both situations. For each source in the \magpis data, we report the median of $P(F|D)$ as the 6\,cm flux estimate, since the true flux lies above or below
this value with equal probability. For strong detections, this value converges to D. For non-detections, the data provide upper limit constraints more
effectively than lower limits, and the median of $P(F|D)$ is comparable to $\sigma$.

These fluxes are given in Table \ref{tab:lyc} for the 40 bubbles covered by the \magpis survey. \citepy{Condon06} derive a relationship between
this flux and the ionizing photon rate:
\begin{align}
\label{eq:nlyc}
\nonumber \biggl( { N_{\rm Ly} \over {\rm s}^{-1} } \biggr) \approx & 6.3 \times 10^{52} 
\biggl( { T_{\rm e} \over 10^4 {\rm ~K} } \biggr)^{-0.45}\\
&\biggl( { \nu\over {\rm GHz} } \biggr)^{0.1} 
\biggl( { L_\nu \over 10^{20} {\rm~W~Hz}^{-1} } \biggr)
\end{align}
this rate is provided in Table \ref{tab:lyc}, along with the number of O9.5 stars needed to produce such a flux \citep{Martins05}. 
These flux estimates are lower limits, since any ionizing photons that are absorbed by dust or escape out of the bubble fronts or backs
are unaccounted for in Equation \ref{eq:nlyc}. \citepy{Watson08} identified the likely stars powering three bubbles, 
and estimated that the \magpis estimate of N$_{\rm Ly}$ is about a factor of two lower than the expected ionizing photon flux from these stars.
Thus, the values listed in Table \ref{tab:lyc} should be considered factor of $\sim 2$ underestimates of the true
ionizing luminosity.

\begin{deluxetable}{lrrr@{}l}
\tablecolumns{5}
\tablewidth{0pc}
\tabletypesize{\scriptsize}
\tablecaption{H {\sc ii} region properties
\label{tab:lyc}}
\tablehead{
\colhead{Bubble} &  \colhead{20\,cm Flux} & \colhead{log N$_{Ly}$} & \colhead{N(O9.5)}\\
\colhead{} & \colhead{(Jy)} & \colhead{(log s$^{-1}$)} & \colhead{}}
\startdata

 N5 & 1.25e$-$01 & 47.13 & &.4 \\ %3.69e$-$01\\
N10 & 4.13e$+$00 & 48.73 & 15& \\ %1.49e$+$01\\
N14 & 2.41e$+$00 & 48.36 & 6&.4 \\ %6.36e$+$00\\
N15 & 1.97e$-$02 & 45.75 & &.02 \\ %1.53e$-$02\\
N16 & 1.73e$-$03 & 44.89 & &.002 \\ %2.14e$-$03\\
N20 & 9.15e$-$02 & 46.87 & &.20 \\ %2.02e$-$01\\
N21 & 2.95e$+$00 & 48.43 & 7&.3 \\ %7.35e$+$00\\
N22 & 3.49e$+$00 & 48.55 & 9&.8 \\ %9.75e$+$00\\
N27 & 5.04e$-$02 & 46.76 & &.16 \\ %1.57e$-$01\\
N29 & 4.34e$+$00 & 48.33 & 5&.8 \\ %5.84e$+$00\\
N30 & 3.51e$+$00 & 48.27 & 5&.1 \\ %5.11e$+$00\\
N34 & 2.37e$-$01 & 47.70 & 1&.4 \\ %1.38e$+$00\\
N35 & 7.57e$+$00 & 49.28 & 53& \\ %5.30e$+$01\\
N36 & 9.15e$+$00 & 49.34 & 60& \\ %5.96e$+$01\\
N37 & 7.04e$-$01 & 47.57 & 1&.0 \\ %1.03e$+$00\\
N39 & 1.60e$+$01 & 49.24 & 47& \\ %4.73e$+$01\\
N40 & 5.19e$-$01 & 47.65 & 1&.2 \\ %1.22e$+$00\\
N44 & 8.80e$-$02 & 47.10 & &.4 \\ %3.50e$-$01\\
N45 & 1.42e$+$00 & 48.33 & 5&.9 \\ %5.92e$+$00\\
N46 & 1.09e$+$00 & 48.29 & 5&.4 \\ %5.41e$+$00\\
N47 & 1.80e$+$00 & 48.56 & 10& \\ %1.01e$+$01\\
N49 & 9.85e$-$01 & 48.21 & 4&.5 \\ %4.49e$+$00\\
N50 & 6.20e$-$01 & 47.85 & 1&.9 \\ %1.93e$+$00\\
N51 & 2.98e$-$02 & 46.75 & &.15 \\ %1.54e$-$01\\
N52 & 7.14e$+$01 & 50.11 & 350& \\ %3.55e$+$02\\
N53 & 4.71e$-$01 & 47.33 & &.58 \\ %5.85e$-$01\\
N54 & 9.29e$-$01 & 48.29 & 5&.4 \\ %5.41e$+$00\\
N56 & 1.46e$-$01 & 47.49 & &.85 \\ %8.49e$-$01\\
N61 & 3.62e$-$01 & 47.43 & &.75 \\ %7.50e$-$01\\
N62 & 5.87e$-$02 & 46.64 & &.12 \\ %1.22e$-$01\\
N65 & 5.76e$-$02 & 46.58 & &.10 \\ %1.04e$-$01\\
N74 & 2.56e$-$02 & 46.03 & &.03 \\ %2.92e$-$02\\
N77 & 1.68e$-$02 & 46.30 & &.06 \\ %5.51e$-$02\\
N79 & 3.94e$-$01 & 46.18 & &.04 \\ %4.16e$-$02\\
N80 & 2.54e$-$01 & 46.21 & &.04 \\ %4.43e$-$02\\
N82 & 1.39e$+$00 & 48.20 & 4&.3 \\ %4.33e$+$00\\
N84 & 5.20e$-$03 & 45.70 & &.01 \\ %1.37e$-$02\\
N88\tablenotemark{a} & 1.26e$+$01 & \nodata & \nodata \\
N90 & 1.61e$-$01 & 47.33 & &.58 \\ %5.83e$-$01\\
N94 & 1.19e$-$02 & 45.83 & &.02 \\ %1.86e$-$02\\
\\
Orion\tablenotemark{b} & \nodata & 48.85 & 20& \\
NGC 3603\tablenotemark{b} & \nodata & 51.05 & 3100&  \\
\enddata
\tablenotetext{1}{No distance information (not detected in CO data)}
\tablenotetext{2}{Data from \citepy{Kennicutt84}}
%\tablenotetext{3}{Outside of {\sc magpis} survey region}
\end{deluxetable}

The \hii regions in this sample are substantially weaker than classical \hii regions. For comparison, we include
data on the Orion Nebula (considered faint by classical \hii standards) and NGC 3603 (a giant \hii region, \citealt{Kennicutt84}). With a few exceptions, most of the free-free emission in this sample is fainter then even the Orion \hii region. 
Most of the bubbles are powered by early B stars, and can 
be viewed as lower-mass equivalents to classical \hii regions. This confirms
Churchwell et al's (2006) original interpretation that bubbles surround late
O and early B stars. However, we note that many of these
objects do have some detectable level of free-free emission, and as such may be driven
by the expansion of the \hii region as opposed to stellar winds \citep{Freyer03}. 

\subsection{Properties of the dense gas}
\label{sec:lvg}
We mapped the HCO$^+$ 4--3 line toward 7 of the 45 bubbles to constrain
the density of the gas in the bubbles.
As the frequency of this line, 357 GHz, is very similar to that
of CO 3--2, maps of each line have similar resolutions ($16''$) and can be directly compared.

HCO$^+$ was detected ($T_{\rm peak}\geq 1$\,K)
in 6 of the 7 bubbles. Its detection indicates high density gas
$n_{\rm H_2}\gtrsim 5\times 10^5$\,cm$^{-3}$ \citep{Evans99}.
However, it is only seen in a single,
generally small, part of the molecular ring where the CO emission
was bright, $T_{\rm peak}\gtrsim 20$\,K. In several cases
these dense regions coincide with CO outflows and, in one case (N74), a YSO
overdensity, indicative of recent star formation (see also $\S\ref{sec:yso}$).
Table \ref{tab:shellprop} lists the peak temperatures of the CO 3--2 and HCO$^+$ 4--3
transitions. Except for bubble 45, which was not detected, there are
two entries for each bubble showing the values where the
integrated intensity of HCO$^+$ is greater than 1\,K\,\kms\
and averaged over the entire ring as defined in $\S\ref{sec:RingProp}$.

We compared these observations with radiative transfer models via
the Large Velocity Gradient formalism using the program LVG in the
MIRIAD\footnote{http://carma.astro.umd.edu/miriad/} package.
For a given gas kinetic temperature,
this program calculates the line radiation temperature of a specified
transition as a function of the molecular column density (normalized by the
linewidth) and H$_2$ volume density.
We created model grids of CO 3--2 and HCO$^+$ 4--3
over the density range $n({\rm H_2})=10^5-10^{7.5}\,{\rm cm}^{-3}$
and normalized column density
$N({\rm CO})/\Delta v=10^{15}-10^{17}\,{\rm cm}^{-2}({\rm km~s^{-1}})^{-1}$
and
$N({\rm HCO^+})/\Delta v=10^{11}-10^{13}\,{\rm cm}^{-2}({\rm km~s^{-1}})^{-1}$.
At these high densities, the 3--2 level of CO is in collisional equilibrium
and the line strength is essentially independent of $n({\rm H_2})$. 
It is mainly dependent on the normalized column density, $N_{\rm CO}/\Delta v$.
As the 4--3 level of HCO$^+$ has a higher critical density,
however, its line strength can be used as a diagnostic of $n({\rm H_2})$
in this range.

The models require kinetic temperatures $T_{\rm kin}\geq 40$\,K 
to match the brighter bubbles (especially N29 and N82).
These high values are not surprising given that the gas lies
on the edge of a wind-swept ionized region around massive stars.
We further find that the inferred values of column and volume density 
are relatively insensitive to $T_{\rm kin}$ in the range $40-60$\,K.

\begin{figure}
\includegraphics[height=3.2in, angle=90]{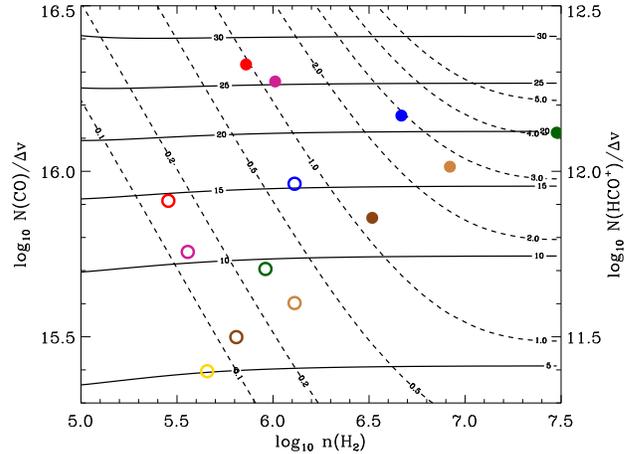}
\caption{LVG model plots of CO and HCO$^+$ peak temperature as a function of column and number density. A kinetic temperature of $T_{\rm k} = 50$K is assumed,
along with abundances [CO]/[H$_2$] = 10$^{-4}$ and [HCO$^+$]/[H$_2$] = 10$^{-8}$.
 The solid lines trace contours of peak CO line intensity, while the dashed lines trace contours of peak HCO$^+$ intensity.
The filled circles correspond to regions of bright HCO$^+$ emission, whereas the open circles correspond to the average over the entire bubble. 
The open circles are upper limits to the HCO$^+$ peak temperature. 
Figure \ref{fig:bubbles}
gives which colors correspond to which bubbles.}
\label{fig:lvg}
\end{figure}

\begin{figure}
\includegraphics[width=3.2in, angle=00]{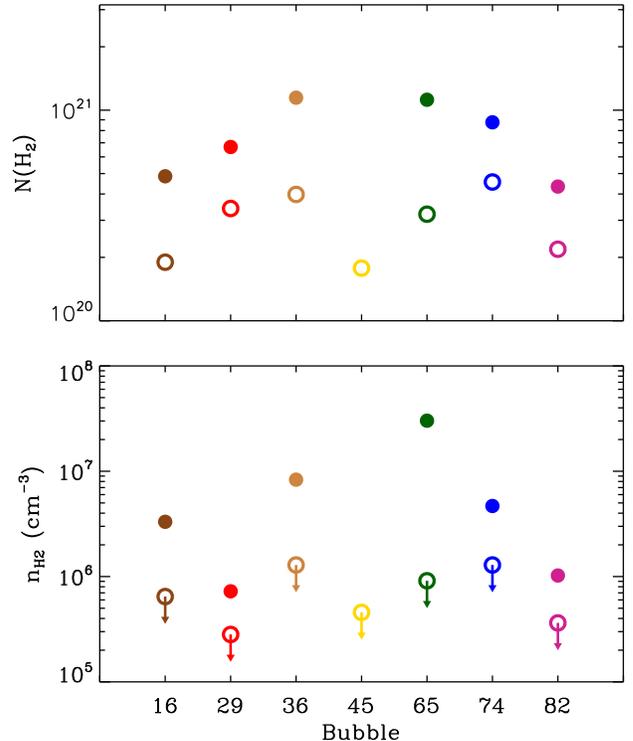}
\caption{The number and column densities for H$_2$, inferred from LVG modeling. As in Figure \ref{fig:lvg}, the filled circles are averages
over regions of bright HCO$^+$ emission, whereas open circles are averages over the entire shell. The open circles in the lower plot are upper limits, since
HCO$^+$ is not detected in the average shell spectrum, and CO emission does not constrain the density.}
\label{fig:bubbles}
\end{figure}

We plot the observed CO and HCO$^+$ line temperatures
on the LVG model grid for $T_{\rm kin}=50$\,K in Figure \ref{fig:lvg}.
The filled circles show the regions of strong HCO$^+$ emission
and the open circles show the averages over the rest of the bubble.
The two are clearly offset indicating, not surprisingly,
both lower column and volume densities away from the HCO$+$ emission.
We multiply the inferred values of $N({\rm CO})/\Delta v$
by the observed CO linewidth and divide by an abundance
${\rm [CO]/[H_2]}=10^{-4}$
to derive the column density of the gas in the bubbles.
The inferred volume density depends on the relative abundance
of HCO$^+$ to CO. Figure \ref{fig:lvg} plots the models for ${\rm [HCO^+]/[CO]}=10^{-4}$
but a lower abundance would shift the HCO$^+$ (dashed) lines up
relative to the CO (solid) lines. In this case, the data points would
move to the right implying higher H$_2$ volume densities.
The CO abundance is fairly well determined since it self-shields
against dissociating radiation and is not depleted in the warm
gas that we are observing. The HCO$^+$ abundance is less certain,
however, especially as the bubbles are effectively large
photo-dissociation regions.
Nevertheless, the relative column densities of HCO$^+$ (e.g., 
comparing two different bubbles, or comparing a region
of bright HCO$^+$ emission to the average over the shell) are still 
well characterized. The values inferred from Figure \ref{fig:lvg} are listed
in Table \ref{tab:shellprop} and plotted in Figure \ref{fig:bubbles}.

\begin{deluxetable*}{llrrrrrr}
\tablecolumns{8}
\tablewidth{0pc}
\tabletypesize{\scriptsize}
\tablecaption{Properties of the shell emission. \label{tab:shellprop}}
\tablehead{
\colhead{Bubble} &  \colhead{Region\tablenotemark{a}} & \colhead{T$^{\rm peak}_{\rm CO}$} & \colhead{$\Delta v_{\rm CO}$} & 
                                       \colhead{T$^{\rm peak}_{\rm HCO^+}$} & \colhead{$\Delta v_{\rm HCO^+}$}  & \colhead{log $n_{\rm H2}$} & \colhead{log N(H2)} \\
\colhead{} & \colhead{} & \colhead{(K)} & \colhead{(km s$^{-1}$)} & \colhead{(K)} & \colhead{(km s$^{-1}$)} & \colhead{(log cm$^{-3}$)} & \colhead{(log cm$^{-2}$)}
}
\startdata
N16 & C & 5.98 & 6.04 & & & 5.81 & 20.28 \\
N16 & H & 12.7 & 6.68 & 1.20 & 2.96 & 6.52 & 20.69 \\
N29 & C & 14.0 & 4.23 & & &  5.45 & 20.53 \\
N29 & H & 27.7 & 3.22 & 1.11 & 2.51 &  5.86 & 20.83 \\
N36 & C & 7.98 & 10.0 & &  & 6.11 & 20.60 \\
N36 & H & 16.8 & 10.0  & 1.59 & 8.57 & 6.92 & 21.06 \\
N45 & C & 4.84 & 7.08 & &  & 5.66 & 20.25 \\
N65 & C & 10.7 & 6.13 & & & 5.96 & 20.51 \\
N65 & H & 16.5 & 8.63 & 3.24 & 6.98 & 7.48 & 21.05 \\
N74 & C & 15.3 & 5.00 & &  & 6.11 & 20.66 \\
N74 & H & 20.9 & 5.92 & 2.39 & 4.32 & 6.67 & 20.94 \\
N82 & C & 11.2 & 3.77 & & & 5.56 & 20.34 \\
N82 & H & 24.4 & 2.33 & 0.65 & 2.75 & 6.01 & 20.64 \\
\enddata
\tablenotetext{1}{H indicates that the flux was averaged over the region of bright HCO$^+$ emission. C indicates that
the flux was averaged over the entire shell, defined by bright CO emission as outlined in $\S\ref{sec:RingProp}$.}
\end{deluxetable*}

We find column densities $N({\rm H_2})=5-10\times 10^{20}\,{\rm cm}^{-2}$
averaged over the regions with HCO$^+$ emission. These are higher than
the average over the rest of the bubble by a factor $\sim 3-7$.
The volume densities, $n({\rm H_2})$ range from $5\times 10^6$ to
$5\times 10^7\,{\rm cm}^{-3}$ and are enhanced by about an order
of magnitude relative to the average in the rest of the bubble.

In exploring the conditions necessary for gravitational collapse, 
\citepy{Whitworth94} show that, in a wide variety of scenarios
(cloud-cloud collisions, supernova remnant expansion,
wind-blown bubbles, and \hii regions),
hydrogen column densities of $4-8 \times 10^{21}\, {\rm cm}^{-2}$ 
are required for layers of molecular gas to collapse. While some of the
dense knots of bright HCO$^+$ in our sample may be unstable to 
gravitational collapse,
the shells on average are an order of magnitude too tenuous to 
collapse due to gravitational instabilities.

\subsection{Star Formation}
As they sweep up and potentially compress ambient material,
the expanding shock fronts from interstellar rings may drive
new epochs of star formation. Here we examine the evidence for triggered 
star formation provided by outflows and overdensities of Young Stellar Objects
(YSOs).

\subsubsection{Young Stellar Objects}
\label{sec:yso}
With photometry alone, it is difficult to determine whether a given object
is embedded in a bubble shell or merely aligned along the line of sight. 
Because of this, associated YSOs must be identified in a statistical sense
by looking for YSO overdensities due to regions of enhanced star formation.

The \glimpse Highly Reliable point source catalogs (GLMIC) provide IRAC colors of 
point sources toward the bubbles. We have used this catalog to identify 
YSOs towards the bubbles in our sample.
To identify likely YSOs, we have used the series of color cuts described
in \citepy{Gutermuth08}. These cuts attempt to discern between bona-fide
YSOs, star forming galaxies and AGN, and non-stellar PAH emission. We apply the cuts
on de-reddened magnitudes; reddening is estimated using the
2MASS-derived all sky extinction maps from \citepy{Rowles09}. 
We then examined each image for clusters of likely YSOs which coincide with bubble features. 

\begin{figure}
\centering
\includegraphics[width= 3.5in]{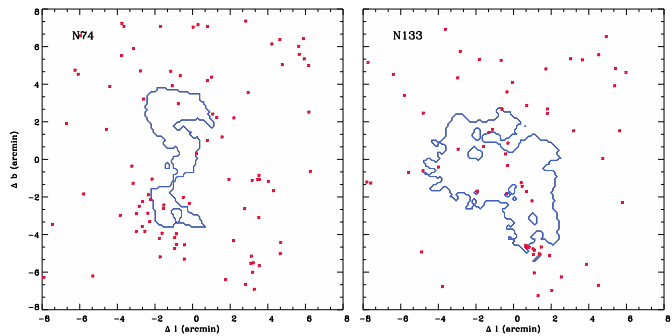}
\caption{
The two bubbles in our sample with YSO overdensities on the shell periphery. The positions of sources flagged as likely YSOs are
shown as points. N74 has an extended cluster of
YSOs at its southeast boundary, while N133 has a tight grouping of YSOs on its southern shell. The contours trace 
the bubble boundaries as discussed in Section \ref{sec:RingProp}}.
\label{fig:yso}
\end{figure}

We find two clear examples of YSO overdensities, which we present in Figure \ref{fig:yso}.
The majority of bubbles have a few likely YSOs on their shells, but at a level comparable
to the field YSO surface density. Six bubbles shells (N15, N21, N27, N34, N39, N44) 
do not overlap any likely YSOs. 

In total, this method yields 109 YSOs which overlap a bubble shell as defined in
$\S$\ref{sec:RingProp}. This represents a moderate enhancement over that expected from
the mean YSO surface density towards these regions (which predicts 55 chance overlaps).

\subsubsection{Outflows}
Young protostars have strong outflows that are often apparent as broad
CO line wings. We searched the moment maps for such features in the
molecular rings and found 12 candidate objects.  Their position and
velocity range are listed in Table~\ref{tab:outflow}.  Since this is
an eye-based identification, the list is biased towards outflows that
are more easily seen on top of the extended shell emission (e.g.,
strong outflows oriented towards our line of sight). At the $\sim
0.25$\,pc resolution of our data, we did not resolve any clear bipolar
structures. The outflow associated with N74 also coincides with the YSO
overdensity there.

\begin{deluxetable}{llrrrr}
\tablecolumns{6}
\tablewidth{0pc}
\tabletypesize{\scriptsize}
\tablecaption{CO outflows\label{tab:outflow}}
\tablehead{
\colhead{Outflow} & \colhead{Bubble} &  \colhead{$l$}  &  \colhead{$b$}  & \colhead{$v_{\rm low}$} & \colhead{$v_{\rm high}$} \\
\colhead{} & \colhead{} & \colhead{($^\circ$)} & \colhead{($^\circ$)} & \colhead{(km\,s$^{-1}$)} & \colhead{(km\,s$^{-1}$)}}
\startdata
1    & N5  & 12.405&    -1.132&     25&    47 \\
2    & N5  & 12.427&    -1.112&     35&    45 \\
3    & N10 & 13.177&     0.061&     40&    50 \\
4    & N20 & 17.932&    -0.664&     34&    46 \\
5    & N34 & 24.295&    -0.153&     85&    115 \\
6    & N36 & 24.850&     0.084&    103&    116 \\
7    & N36 & 24.788&     0.082&     90&    135 \\
8    & N37 & 25.285&     0.266&     30&    40 \\
9    & N40 & 25.389&    -0.368&     30&    54 \\
10   & N49 & 28.830&    -0.254&     80&    100 \\
11   & N65 & 35.020&     0.348&     45&    70 \\
12   & N74 & 38.934&    -0.362&     18&    56 \\
\enddata
\end{deluxetable}

\subsubsection{Extended Green Objects}
\citepy{Cyganowski08} recently identified several hundred Extended Green Objects (or EGOs)
in the \glimpse images. These objects are resolved structures which emit brightly at
$4.5 \mu$m, and are believed to trace shocked molecular gas due to outflows from
massive young stars. Three of these objects overlap the molecular shells of bubbles
N39, N49, and N65, and two of these three (those associated with N49 and N65) are also detected as outflows in CO.

\section{Discussion and Comparison with Previous Work}
\label{sec:discussion} 

\subsection{Rings and Molecular Clouds}

The ring morphology of the CO emission implies that the molecular clouds in which bubbles 
form are flattened, with thicknesses not greater than the bubble sizes of a few pc. Inferring
the three dimensional structure of molecular clouds is a difficult endeavor, and previous
efforts have yielded mixed results. \citepy{Kim08} use supernova light echoes to probe the 
structure of the ISM behind Cas A. They observe numerous filamentary and sheetlike structures,
with a low filling factor of $\sim$0.4\%. However, their technique filters out spatial structures
larger than $\sim$2 pc, and their data probe only the substructures in a potentially
more extended region. 

Statistically analyzing a set of $10^4$ clouds, \citepy{Kerton03} 
compare the observed distribution of molecular cloud axis ratios to those
expected from triaxial ellipsoids, and conclude -- in conflict with our result -- that molecular clouds
are prolate. The results of this technique, however, are sensitive to the choice of model
figures (triaxial ellipsoids in Kerton et al.'s analysis). For example, the same technique has been applied to 
studies of molecular cores. Using two different classes of shapes (axisymmetric vs. triaxial ellipsoids), 
\citepy{Ryden96} and \citepy{Jones02} come to opposite conclusions about whether molecular cores
are prolate or oblate. The flattened nature of molecular clouds implied by our data is independent of
these kinds of model assumptions, and is a direct inference from the observed lack of bubble fronts and backs.

Our CO maps were limited to the bubbles and their environs. However, in many cases, we see extended emission
to the edges of our maps. In general, the bubbles lie in moderate-sized molecular clouds with dimensions
greater than 10 pc in the plane of the ring. That is, the clouds have aspect ratios of at least a few, and
perhaps as much as 10. 

Sheet-like ISM morphologies may naturally arise in molecular clouds created
by the turbulent collision and compression of warm neutral flows in
the Galaxy (see, e.g., numerical work by \citealt{Heitsch05},
\citealt{Vazquez06}, and references therein). The pervasive
oblateness of the bubble-hosting clouds may be relevant to studies of
molecular cloud creation and, by extension, star formation 
(see recent reviews by \citealt{McKee07} and \citealt{Zinnecker07}).

\subsection{Rings and Triggered Star Formation}
While studies of individual ``triggered'' or ``induced'' star formation regions abound in
the literature (recent papers on triggered star formation around bubbles and \hii regions
include \citealt{Deharveng09}, \citealt{Snider09}, \citealt{Koenig08}, and \citealt{Pomares09}), there is less 
consensus on the extent to which the varied triggering mechanisms contribute to the
net stellar production in our Galaxy.

The observation that molecular gas lies in a thin ring around bubble-blowing stars may well impact
the ability for these bubbles to induce new generations of star formation. In such a configuration,
\hii regions likely lose pressure -- and their ability to compress molecular gas -- upon breaking out
of the cloud. Similarly, momentum driven winds encounter a much smaller solid angle of molecular gas
to sweep up. We have cataloged
four tracers of future or ongoing star formation activity towards these regions: dense gas revealed by
HCO$^+$ emission, CO outflows, overdensities of Young Stellar Objects, and coincident Extended
Green Objects. Here we examine to what extent these data constrain estimates of triggered star formation activity.

The low detected fraction of dense HCO$^+$ gas suggests that bubble shells are unstable
to gravitational collapse only in small regions; the shells as a whole do not seem to be 
undergoing gravitational collapse. 

Because bubbles are powered by early B and late O stars (M $\gtrsim\, 10 M_\odot$), 
any self-sustaining triggered star formation in these regions must produce stars
massive enough to blow their own bubbles.  Assuming a typical IMF (e.g. \citealt{Muench02}),
an equivalent statement is that bubble-induced star formation must produce $\gtrsim 250$ 
new stars in order to form a star massive enough drive a secondary bubble.

The completeness of the YSO sample extracted from the \glimpse
data is limited by the bright nebular background. Studying M17, \citepy{Povich09} found that the \glimpse
Point Source Archive recovers essentially all YSOs above $\sim 3 M_\odot$ towards these regions, 
while the completeness falls to $\sim 10\%$ at $M = 2 M_\odot$, and $\sim 1\%$ at $M = 1 M_\odot$.
Our sample is less complete, as we have restricted our focus to the smaller Highly Reliable
Catalog, and most of our bubbles are 1.5--3 times more distant than M17.

Two bubbles show suggestive YSO overdensities, consisting of roughly 10 and 20 associated objects. 
If we assume, for the sake of simplicity, that no objects less massive than 3 M$_\odot$ are detected, then each
flagged YSO represents about 1 in 50 total stars formed via a \citepy{Muench02} IMF. 
Hence, it is possible that these overdensities trace clusters of $\gtrsim 500-1000$ members, 
which would be sufficiently large to create new generations of stellar bubbles. 

The handful of EGOs and CO outflows found towards these bubbles further suggests some degree
of ongoing star formation in these regions. Unfortunately, it is difficult to draw conclusions
about the extent of star formation suggested by these objects. Molecular outflows are a feature of both
high and low mass star formation, and do not constrain the number or mass of stars being formed. 
While EGOs are believed to trace specifically massive star formation, there are too
few detections to draw any statistically meaningful conclusion; the overlap of EGOs and
bubble shells is higher than that expected from chance alignments (3 alignments compared to .35 expected),
but the Poisson uncertainties on this result are very large.

In all, up to 12 of the 43 bubbles in our sample may be undergoing current star formation
as suggested by the above evidence. \citepy{Churchwell07} also looked for triggered star formation 
around bubbles (by looking for secondary bubbles, and YSO coincidences), 
and found evidence of such around 12\% of the objects. 

Identifying triggered YSOs inside bubble shells is complicated by the fact
that these objects are smaller and more distant than more heavily studied regions
\citep{Koenig08, Povich09}. Furthermore, it is difficult to confirm that a given
YSO was formed as the result of a triggering event as opposed to forming independently nearby. 
The soon-to-be-operational SCUBA2
instrument on the \jcmt will circumvent the limitations inherent to this method by providing sensitive
measurements of the cold gas and dust column density towards these bubbles. These data will better determine
whether or not these rings are gravitationally unstable, and hence whether the physical
conditions of interstellar bubbles are conducive to star formation.

\section{Conclusions} 
\label{sec:conclusion}
We have presented new CO 3--2 maps of 43 Spitzer identified
bubbles in the Galactic plane.  These maps point towards a new
interpretation for bubble morphology - namely, that the shock fronts
driven by massive stars tend to create rings instead of
spherical shells. We suggest that this morphology naturally arises
because the host molecular clouds are oblate, even sheet-like, with thicknesses of a few pc, and
widths $\gtrsim10$\,pc.
Such a morphology implies that expanding shock fronts are poorly bound by
molecular gas, and that cloud compression by these shocks may be limited. This 
idea is borne out by observations in HCO$^+$, which show dense gas to be confined to 
small regions of bubble shells.  

We thank Ed Churchwell for stimulating discussions and the NSF for support through grant AST-0808144.

%\begin{thebibliography}

%\bibliography{bib}
%\bibliographystyle{apj}

\end{document}